\documentclass[aip,jcp,reprint,amssymb,amsmath,amsfonts,superscriptaddress,showpacs,floatfix]{revtex4-1} % JCP for checking your page length
\usepackage{hyperref}
\hypersetup{colorlinks,citecolor=blue,filecolor=black,linkcolor=blue,urlcolor=black}

\usepackage{graphicx} %amsmath,amssymb,amsfonts already included in document class?!

\usepackage{lmodern} %to get rid of subsubsection font problems
\usepackage[caption=false]{subfig}
\usepackage{siunitx}
\usepackage{bm} %for bold math symbols and greek letters
\usepackage{color}
\usepackage[usenames,dvipsnames]{xcolor}
\usepackage{upgreek}
\usepackage{natbib}
\usepackage[normalem]{ulem}

\definecolor{ppurple}{RGB}{98,111,179}
\definecolor{porange}{RGB}{240,145,55}
\definecolor{pgreen}{RGB}{53,178,87}
\definecolor{pred}{RGB}{220,73,89}
\definecolor{pblue}{RGB}{7,174,227}
\definecolor{orange}{rgb}{1,0.5,0}

\newcommand\vecc[1]{{\bf{#1}}}

\newcommand*{\mathcolor}{}
\def\mathcolor#1#{\mathcoloraux{#1}}
\newcommand*{\mathcoloraux}[3]{%
  \protect\leavevmode
  \begingroup
    \color#1{#2}#3%
  \endgroup
}

\let\OldBlacksquare\blacksquare \renewcommand{\blacksquare}{{\mathcolor{gray}{\OldBlacksquare}}}

\usepackage{textcomp}

\usepackage[nointegrals]{wasysym}
\usepackage{pifont}

\usepackage[geometry]{ifsym}

\let\OldBoxplus\boxplus \renewcommand{\boxplus}{{\mathcolor{gray}{\OldBoxplus}}}

\usepackage{cleveref}
\usepackage{tikz}
\usetikzlibrary{shapes.geometric}

\crefname{figure}{Fig.}{Figs.}
\crefname{equation}{Eq.}{Eqs.}
\crefname{section}{Sec.}{Secs.}
\crefname{table}{Tab.}{Tabs.}

\begin{document}
\title{Colloids Exposed to Random Potential Energy Landscapes: from Particle Number Density to Particle-Potential and Particle-Particle Interactions}
\author{J\"org Bewerunge}
\affiliation{Condensed Matter Physics Laboratory, Heinrich Heine University, 40225 D\"usseldorf, Germany}
\author{Ankush Sengupta}
\affiliation{Department of Chemical Physics, Weizmann Institute of Science, 234 Herzl St., Rehovot 7610001, Israel}
\author{Ronja F. Capellmann}
\affiliation{Condensed Matter Physics Laboratory, Heinrich Heine University, 40225 D\"usseldorf, Germany}
\author{Florian Platten}
\affiliation{Condensed Matter Physics Laboratory, Heinrich Heine University, 40225 D\"usseldorf, Germany}
\author{Surajit Sengupta}
\affiliation{TIFR Centre for Interdisciplinary Sciences, Hyderabad 500075, India}
\author{Stefan U. Egelhaaf}
\affiliation{Condensed Matter Physics Laboratory, Heinrich Heine University, 40225 D\"usseldorf, Germany}

\date{\today}

%%%%% Abstract %%%%%

\begin{abstract}
Colloidal particles were exposed to a random potential energy landscape (rPEL) that has been created optically via a speckle pattern.
The mean particle density as well as the potential roughness, i.e. the disorder strength, were varied.
The local probability density of the particles as well as its main characteristics were determined.
For the first time, the disorder-averaged pair density correlation function $g^{(1)}(r)$ and an analogue of the Edwards-Anderson order parameter $g^{(2)}(r)$, which quantifies the correlation of the mean local density among disorder realisations, were measured experimentally and shown to be consistent with replica liquid state theory results.
\end{abstract}

%05.20.Jj Statistical mechanics of classical fluids
%47.57.-s: Complex fluids and colloidal systems; "Fluid Dynamics"
%64.70.pv Colloids (64: Equations of state, phase equilibria, and phase transitions; Condensed Matter)
%82.70.Dd colloids (82: Physical chemistry and chemical physics; Interdisciplinary Physics)

\maketitle

%%%%% Introduction %%%%%

\section{Introduction}
\label{sec:introduction}

The potential energy landscape (PEL) of a system depends on the coordinates and/or other parameters of its constituents.\cite{Wales2004}
The concept of a PEL is successfully used in many fields of science to determine the properties and behavior of systems ranging from small, large and polymeric molecules, proteins and other biomolecules to clusters, glasses and biological cells.\cite{Wales2004}
It is also applied to describe the transport over atomic surfaces,\cite{Barth2000, Jardine2004, Nguyen2014} in materials with defects (e.g., ions in zeolites \cite{Chen2000} or charge carriers in conductors with impurities \cite{Heuer2005}), in inhomogeneous media \cite{Isichenko1992, Scher1973} (e.g., porous gels,\cite{Dickson1996} cell membranes \cite{Hsieh2014} or cells \cite{Weiss2004, Hofling2013, Tolic-Norrelykke2004, Blainey2009}) or in the presence of fixed obstacles as in a Lorentz gas.\cite{Hofling2006}
They are also used to determine the rates of (bio)chemical reactions,\cite{Wales2004, Min2009} the folding of proteins and DNA,\cite{Baldwin1994, Durbin1996, Janovjak2007, Arcella2014, Frauenfelder1991, Dill1985, Bryngelson1989} as well as the particle dynamics in dense suspensions close to freezing,\cite{Indrani1994} in glasses \cite{Kirkpatrick1989, Heuer2008, Gotze1999, Angell1995, Sciortino2005, Goldstein1969, Sastry1998, Nguyen2014, Charbonneau2014, Ediger1996, Stillinger1995, Royall2015} or, more general, in crowded systems.\cite{Hofling2013}

We focus on random potential energy landscapes (rPEL), which have been used in the interpretation of several experimental observations.
For example, rPEL with a Gaussian distribution of energy values with a width of about the thermal energy have been used to describe the behavior of RNA, proteins and transmembrane helices.\cite{Hyeon2003, Janovjak2007, Goychuk2014}
Although a rPEL might only represent a crude approximation for many experimental situations, it often provides a very useful initial description of the effect of disorder on the dynamics.\cite{Chen2000, Wolynes1992, Royall2015}

The PEL is experimentally realised by exploiting the interaction of light with colloidal particles,\cite{Ashkin1986,Ashkin1992} which was already applied to realise, e.g., sinusoidal~\cite{Ackerson1987,Bechinger2001,Jenkins2008b,Dalle-Ferrier2011,Juniper2012} or random landscapes.\cite{Douglass2012,Hanes2012a,Hanes2012b,Evers2013b,Volpe2014,Bewerunge2016a}
(See \onlinecite{Evers2013b} for a review.)
Here we investigate how a rPEL modifies the spatial arrangement of ensembles of colloidal particles.\cite{Sengupta2005, Isichenko1992, Wales2004}
Local density variations occur, which are related to the distribution of energy levels $p(U)$ and the spatial correlation function $C_U(r)$ of the underlying potential.
For various disorder strengths, controlled through the laser power $P$, and particle concentrations, i.e. mean particle number densities $\rho_0$, we track particle positions and calculate the local density $\rho({\bf r},t)$ at each time $t$, based on which different correlation functions are obtained:
the disorder-averaged pair distribution function or pair density correlation function $g^{(1)}(r)$,~\cite{Hansen2006} and, to characterize the quenched disorder, the density correlation $g^{(2)}(r)$,~\cite{Menon1994,Sengupta2005} similar to the Edwards-Anderson order parameter,~\cite{Edwards1975, Sommers1982, Parisi2012} which is intensively used in the context of spin glasses and has been proposed in the context of pinned vortex liquids~\cite{Menon1994} and calculated in computer simulations.~\cite{Sengupta2005,Dasgupta2006}
However, as yet it has never been measured in an experiment.
In this paper, we proceed to do precisely that.
This analysis provides the main characteristics of the effect of the disorder, i.e.~the rPEL, with respect to particle-potential as well as pair and higher order inter-particle interactions and can easily be extended to other systems, such as magnetic bubble arrays in a disordered potential,~\cite{Seshadri1993a,Seshadri1993b,Seshadri1993c} particles on patterned surfaces~\cite{Wu2006} and vortex liquids as well as glasses in the presence of random pinning.~\cite{Dasgupta1998,Dasgupta2006}

%%%%% Section 1 %%%%%
\section{Materials and Methods}
\label{sec:materials}

\subsection{Optical Set-up}
\label{sec:setup}

A random intensity distribution, i.e. a speckle pattern, was created by directing an expanded laser beam (Laser Quantum, Opus 532, wavelength $532$~nm, maximum intensity $P_{\mathrm{max}} = 2.6$~W) onto a microlens array (RPC Photonics, Engineered Diffuser\texttrademark\, EDC-1-A-1r, diameter $25.4$~mm)\cite{Sales2003, Dickey2014} and subsequently focussing the modified beam into the sample plane of an inverted microscope.
This results in a macroscopically uniform beam with a so called top-hat intensity distribution.
However, the wavefronts from the randomly-distributed microlenses interfere in the sample plane.
This leads to microscopic intensity variations, so-called laser speckles, to which the particles were exposed.
The interaction of the particles with the speckle pattern can be described by a rPEL.
The particle size roughly matches the speckle size, but is much larger than the laser wavelength. Moreover, the laser intensity is spread over a large field of view. Thus, we neither expect nor observe optical binding effects~\cite{Burns1989,Bowman2013} or light field-induced dispersion forces.\cite{Brugger2015}
The colloidal particles were observed using the inverted microscope (Nikon, Eclipse Ti-U) with a $20\times$ objective (Nikon, CFI S Plan Fluor ELWD, numerical aperture $0.45$).
A detailed description of the optical set-up and a statistical analysis of both the intensity pattern and the resulting rPEL can be found in ref. \onlinecite{Bewerunge2016a}, where the present conditions correspond to `BE $5\times$'.

\subsection{Samples}
\label{sec:sample}

Samples consisted of spherical polystyrene particles with sulfonated chain ends (Invitrogen, diameter $D=2.8\; \upmu$m, polydispersity 3.2~\%) dispersed in purified water (ELGA purelab flex, electrical resistivity $18.2\times10^4~\Omega\text{m}$).
Three glass cover slips ($\#1.5$) and a microscope slide (all from VWR) were assembled to form a small capillary.\cite{Jenkins2008}
After the capillary was filled with the dispersion, it was sealed with UV-glue (Norland, NOA61).
Due to the density difference between particles and water, the particles sedimented and formed a quasi two-dimensional layer at the bottom of the sample cell.

\subsection{Data Acquisition}
\label{sec:data}

Each measurement consisted of $K\approx27,000$ images, which were recorded at 3.75 frames per second using an 8-bit camera (AVT, Pike F-032B with $640\times480$ pixels and pixel pitch of $0.372$ $\upmu\text{m}$).
Particle positions were determined using standard procedures.\cite{Crocker1996}
Because the system evolves from a quenched random distribution towards its equilibrium distribution, care was taken that the correlation functions are not affected by the relaxation process, i.e. do not show a time dependence.\cite{Bewerunge2016b}

Based on the particle positions, we determined the number of particles $\mathcal{N}(x_m,y_n,t,l)$ in each region at $\vecc{r}=\vecc{r}_{mn}=(x_m,y_n)$ at each time $t$ for a particular realisation of the potential $l$ (out of $L$ different realisations), and calculated the local particle density as
\begin{equation}
\rho(x_m,y_n,t,l)=\frac{\mathcal{N}(x_m,y_n,t,l)}{\Delta x \Delta y} \text{,}
\label{eq:data}
\end{equation}
where $\Delta x=x_m-x_{m-1}$ and $\Delta y = y_m-y_{m-1}$, with $\Delta x = \Delta y$ for all $m=1...M$, $n=1...N$.
Hence the quadratic regions all have the same size of $0.186~\upmu$m, which is well above the uncertainty of the particle positions, about $0.05~\upmu$m.\cite{Crocker1996}
It is noteworthy that these regions do not coincide with pixels of the camera.
The distance $r$ between two regions at $\vecc{r}$ and $\vecc{r}'$ is $r = |\vecc{r}-\vecc{r}'|$, which depends on the location of both regions and thus on $m$, $m'$, $n$ and $n'$.
It was divided into bins of $\Delta r = 0.2 \Delta x$, which represents a compromise between good statistics and high resolution.

%%%%% Results %%%%%
\section{Results and Discussion}
\label{sec:results}

\subsection{Random Potential Energy Landscape (rPEL)}
\label{sec:speckle}

The colloidal particles were exposed to a rPEL by exploiting the interaction of light with particles having a refractive index different from the one of the dispersing liquid.
Their interaction usually is described by two forces:\cite{Ashkin1986,Ashkin1992} a scattering force, which pushes the particles along the beam, and a gradient force, which pulls particles with a larger refractive index than the one of the solvent towards regions of high intensity.
This effect is typically applied in optical tweezers which are used to trap or manipulate colloidal particles.\cite{Ashkin1986,Ashkin1992,Grier2003,Dholakia2008}
Rather than single focused beams, an extended light field can be used to create a PEL.\cite{Evers2013b}
To predict not only the shape of the PEL but also its amplitude, the particles' susceptibility or polarizability needs to be known, which typically is not the case.
Nevertheless, it is possible to calculate the typical characteristics of the PEL by integrating the local intensity $I(\vecc{r})$ over the particle's projected volume, thus taking the particle volume traversed by the light beam into account.\cite{Evers2013a}
This results in an estimate of the potential $U(\vecc{r})$ imposed on a particle, which then is considered to be point-like.\cite{Bewerunge2016a}

\begin{figure} %%%%% 111111 %%%%%
\includegraphics[width=1.0\linewidth]{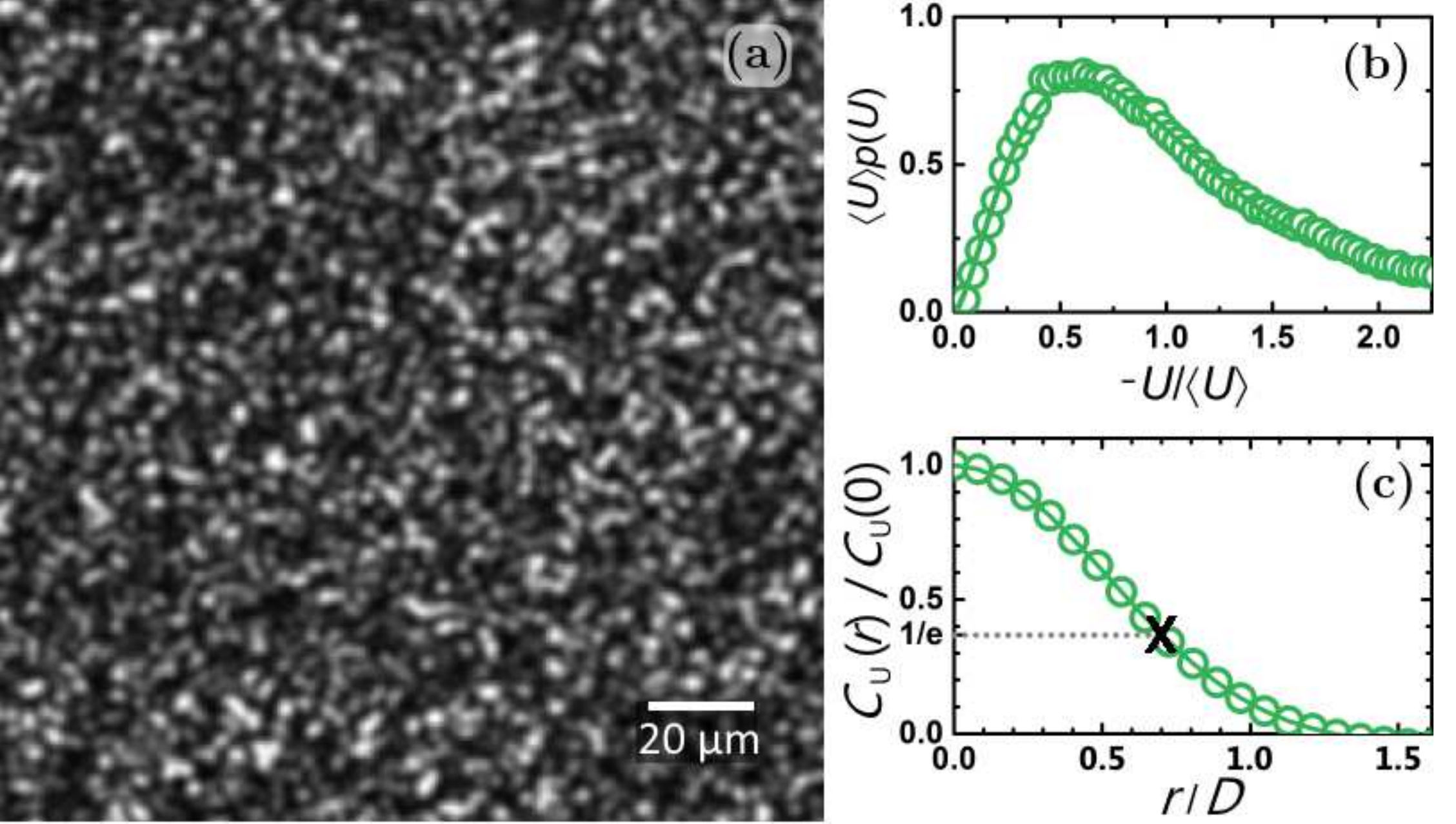}
\caption[Speckle pattern (a), $p(U)$ (b) and $C_{\text{U}}(r)$ (c).]
{(a) Random potential energy landscape (rPEL), i.e. $U(\vecc{r})$, as calculated by convolving the measured intensity pattern $I(\vecc{r})$ with the projected volume of a particle of diameter $D = 2.8~\upmu$m, (b) its normalized probability density of energy values $p(U)$ and (c) its normalized spatial correlation function $C_{\text{U}}(r)/C_{\text{U}}(0)$ with 1/e-width $0.69D$ indicated by a cross.
\label{fig:speckles}}
\end{figure}

\Cref{fig:speckles} (a) shows one realisation of the rPEL, i.e. $U(\vecc{r})$, as a grey scale image which was obtained by convolving a recorded intensity pattern with the projected volume of a particle.
The rPEL was characterized by the distribution of energy values $p(U)$, which follows a Gamma distribution\cite{Goodman2007} with shape parameter $M =2.6$ (\cref{fig:speckles} (b), for details see condition BE5$\times$ in Tab. II of ref.~\onlinecite{Bewerunge2016a}).
The length scale of the fluctuations was described by the normalized spatial covariance function $C_U (\vecc{r}) = \langle U(\vecc{r'})U(\vecc{r'}+\vecc{r})\rangle_{\vecc{r'}} / \langle U(\vecc{r'})\rangle_{\vecc{r}'}^2 - 1$, whose azimuthal average can be described by a Gaussian distribution $C_U(r) = \exp(-(r/\xi)^2)$ with $\xi = 0.69D$ (\cref{fig:speckles} (c)).

\subsection{Particles in the rPEL}
\label{sec:sampling}

\begin{figure} %%%%% 222222 %%%%%
\includegraphics[width=0.9\linewidth]{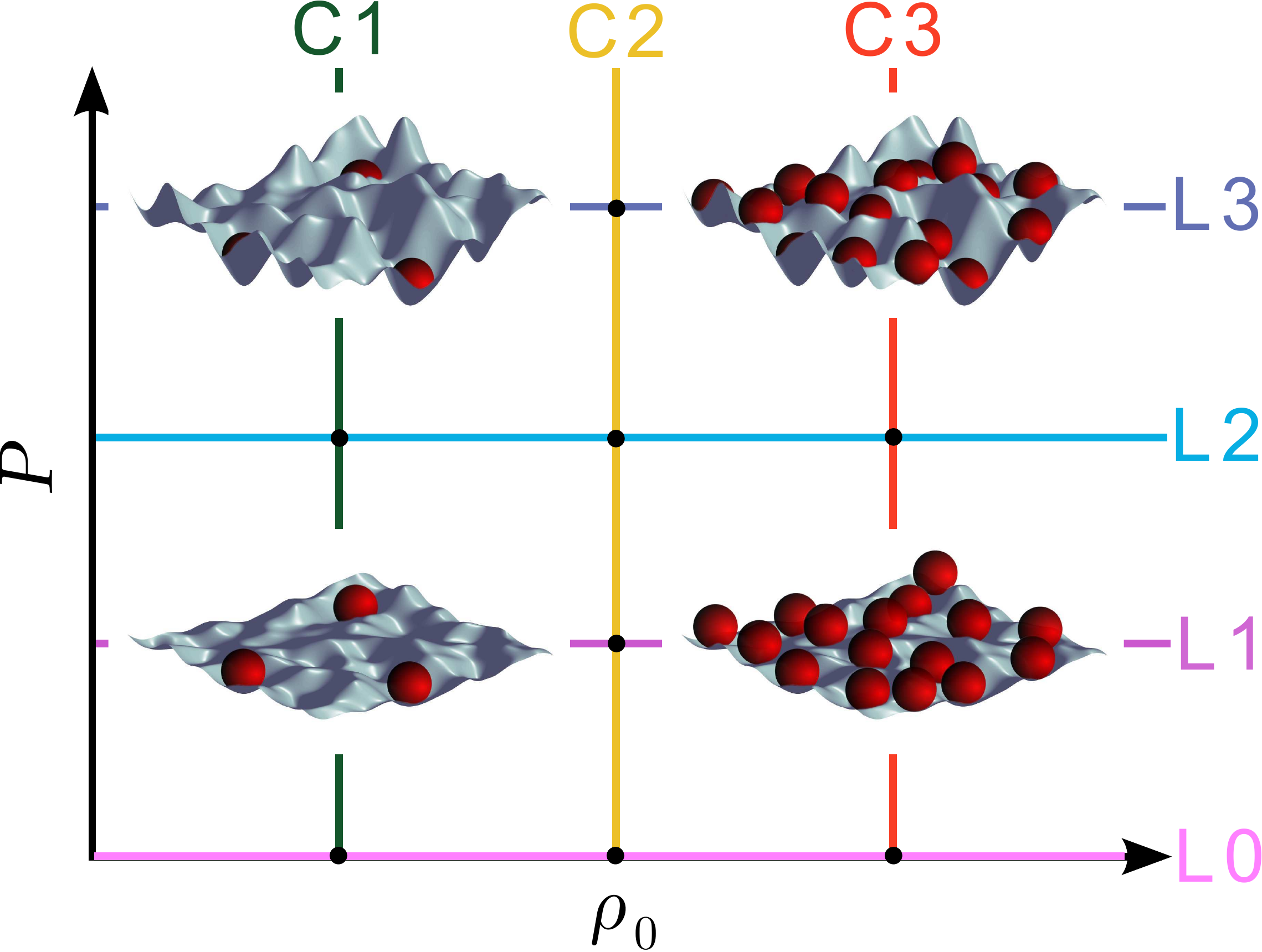}
\caption[$P$ vs. $\rho_0$]
{Different laser powers $P$ (L0 to L3), corresponding to different mean potential values $\langle U\rangle$ or disorder strengths, and mean particle densities $\rho_0$ (C1 to C3) are investigated.
For four conditions sketches showing particles in a rPEL are shown.
\label{fig:PlvsCon}}
\end{figure}

In the experiments, the particle concentration, i.e. the mean particle number density $\rho_0$ or the particle area fraction $\phi_\text{A}=\pi (D/2)^2\rho_0$, as well as the laser power $P$, and hence the mean potential value $\langle U \rangle$ and the disorder strength, were varied, whereas the shape of the distribution, $p(U)$, and the spatial correlation function, $C_U(r)$, remain unchanged (\Cref{fig:PlvsCon}).
We consider three different $\rho_0$ (C1: $\rho_0=0.007~\upmu \text{m}^{-2}$, C2: $\rho_0=0.041~\upmu \text{m}^{-2}$, C3: $\rho_0=0.072~\upmu \text{m}^{-2}$, corresponding to area fractions $\phi_\text{A}= 0.045$, $0.25$ and $0.45$, respectively) as well as four different $P$ (L0: 0 mW, L1: 917 mW, L2: 1640 mW, L3: 2600 mW), and indicate conditions by CiLj.

\begin{figure*} %%%%% 333333 %%%%%
\includegraphics[width=1.0\linewidth]{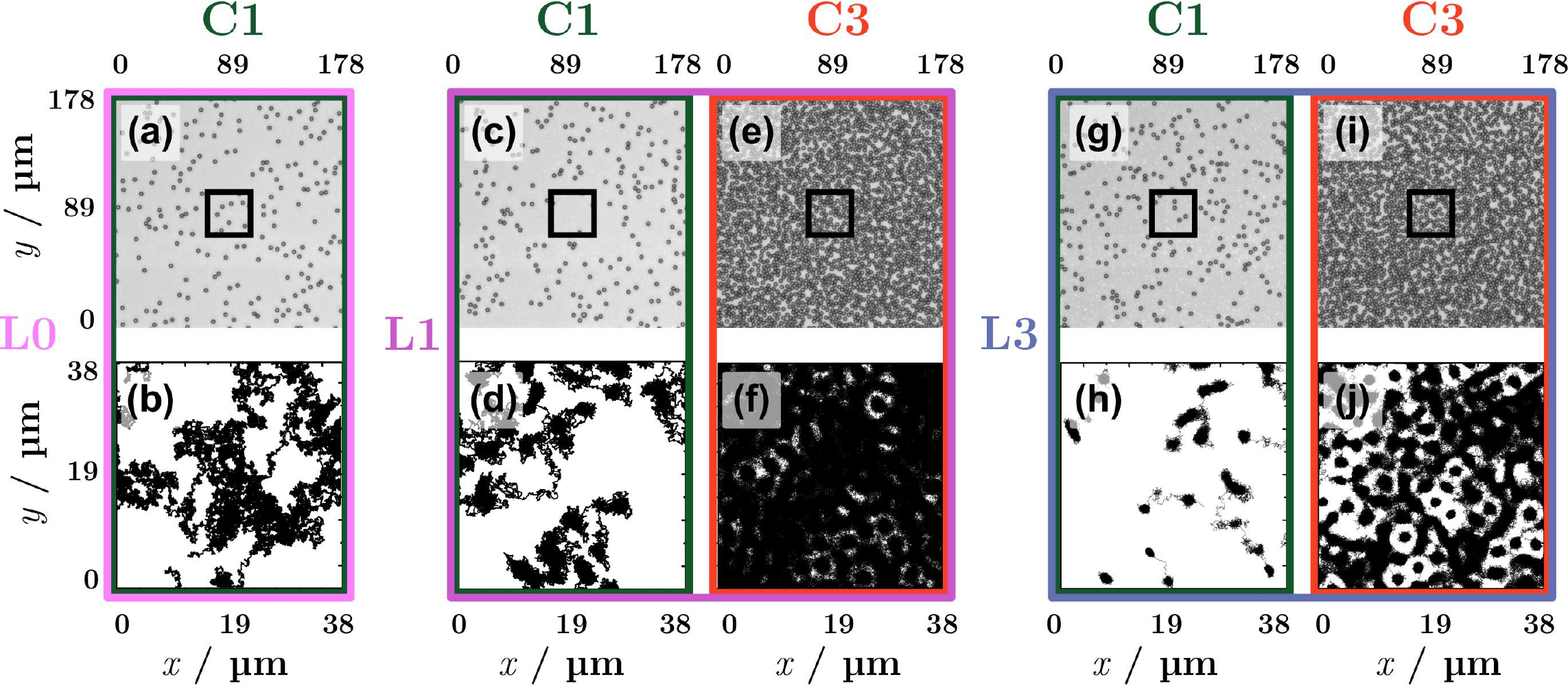}
\caption[images]
{(top) Micrographs of parts of the samples ($178\times178~\upmu\text{m}^2$) and (bottom) particle trajectories in a central region ($38\times38~\upmu\text{m}^2$, indicated in the micrographs) during a time $\Delta t=7200$~s after the micrograph has been taken, for different laser powers L0, L1, L3 (left to right) and mean particle densities C1, C3.
\label{fig:images}}
\end{figure*}

\Cref{fig:images} shows images of colloidal particles (top) and their trajectories (bottom) for two different mean particle densities $\rho_0$ (C1, C3) and increasing laser power $P$ and hence disorder strength (L0, L1, L3), where L0 corresponds to the absence of a laser field and hence free diffusion.
(For images at other combinations of mean particle density and laser power see~\cref{fig:A1} in the appendix.)
Neither for the low nor for the high mean particle density an effect of the potential is immediately visible in the images.
However, there is a clear effect of the rPEL on the trajectories. 
For the low mean particle density C1, as the disorder strength is increased, the motion of the particles is restricted to small areas and a few particles even stay in one potential minimum for the entire measurement time.
At high mean particle density C3 and low laser power L1 (\cref{fig:images} (f)) almost the whole field of view is sampled by the particles.
This indicates that the particles are very mobile and exchange positions.
In contrast, for high potential roughness L3 (\cref{fig:images} (j)) some particles appear stuck in potential minima.
This prevents other particles from exploring their neighbourhood and leads to regions depleted of particle centres.

The dynamic behaviour has important consequences on how particles sample a PEL.
Since experiments have a limited measurement time, sampling can be incomplete and hence local information only be partially accessible.
The completeness of sampling determines whether time-averaged quantities might hold reliable local information and describe all points in a PEL, or whether only spatially-averaged quantities might provide reliable information.
Very low mean particle densities result in only limited information on some locations of the PEL.
Upon increasing the mean particle density, sampling can become more complete (e.g. C3L1).
However, higher mean particle densities also enhance particle-particle interactions, which hence might dominate particle-potential interactions.
This reduces correlations with the underlying potential.
Moreover, a strongly varying potential can also result in an `undersampling' of energetically unfavourable areas, i.e. potential maxima, since they are avoided by the particles.
The unexplored areas might depend on the initial positions of the particles, due to the quenched disorder of the potential.
An average over different disorder realisations might help, but excludes the determination of local quantities, which loose their relevance.

\subsection{Time-Averaged Particle Density}
\label{sec:dense}

First, we consider the time-averaged (or thermal-averaged) local particle density
\begin{equation}
\left \langle \rho \left(\vecc{r}, t, l\right) \right \rangle_t = \frac{1}{K}\sum_{k=1}^{K}\rho(x_m,x_n,t,l)\:\text{.}
\end{equation}
Its ensemble and disorder average gives the mean particle density $\rho_0 = [\left\langle\rho\left(\vecc{r},t,l\right)\right\rangle_{t,\vecc{r}}]_l$,
where $\langle ... \rangle_t$, $\langle ... \rangle_\vecc{r}$ and $[...]_l$ denote time, ensemble and disorder averages, respectively.
In the experiments presented here, the large field of view provides a sufficient disorder average within a single rPEL realisation.
Thus here the total number of disorder realisations $L=1$ and the sample average implies an ensemble and disorder average.

\begin{figure} %%%%% 444444 %%%%%
\includegraphics[width=1.0\linewidth]{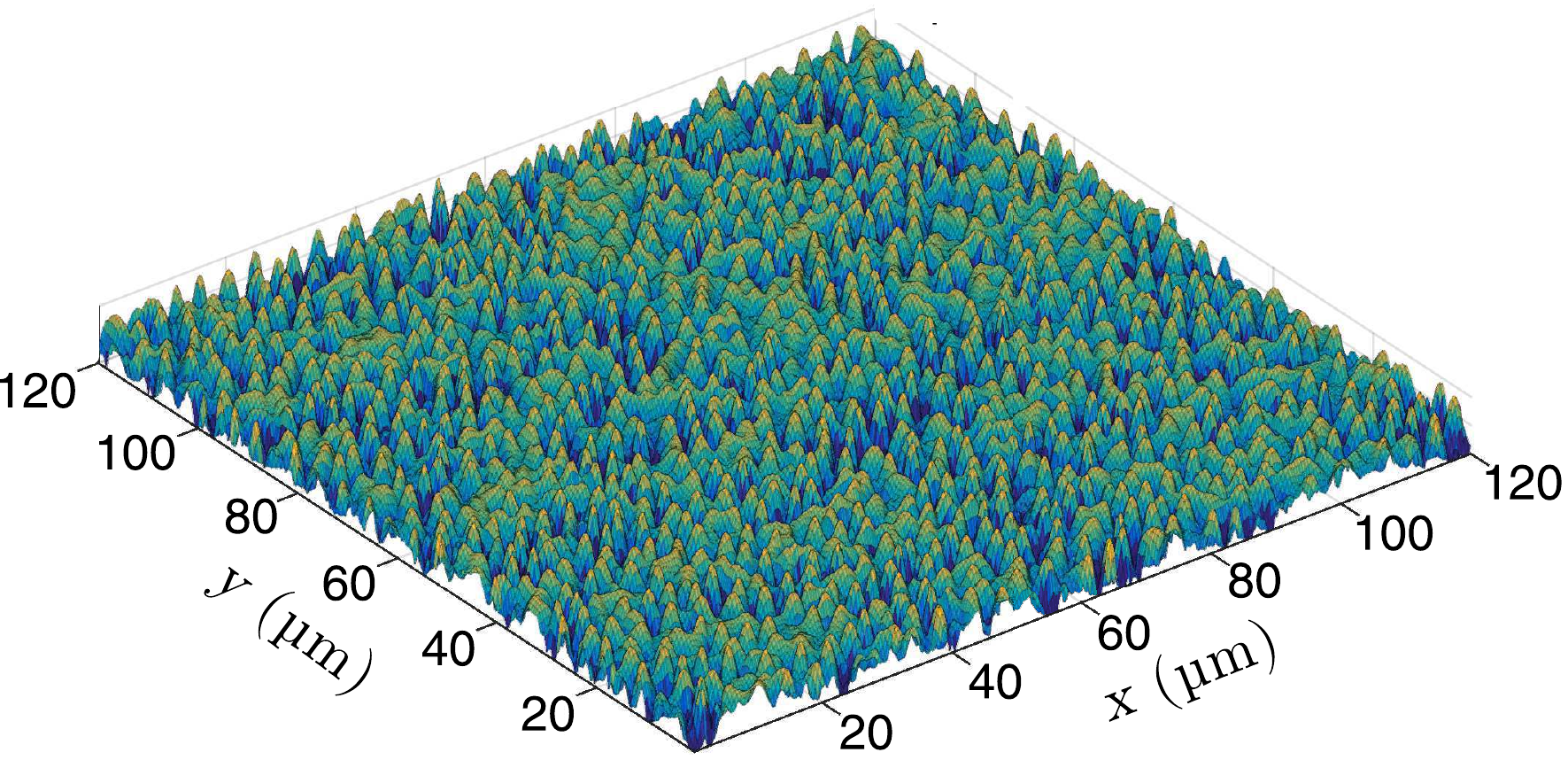}
\caption[rhoxy]
{Time-averaged local particle density $\langle \rho(\vecc{r},t) \rangle_t$ for high laser power L3 and large mean particle density C3.
The logarithmic colour scale indicates low ($\rho_0=1\times10^{-4}~\upmu \text{m}^{-2}$) to high ($\rho_0=0.63~\upmu \text{m}^{-2}$) local densities by dark blue to red colours.
\label{fig:rhoxy}}
\end{figure}

\Cref{fig:rhoxy} shows the time-averaged local particle density $\langle \rho(\vecc{r},t)\rangle_t$, for large laser power L3 and high mean particle density C3 (cf.~\cref{fig:images} (i) and (j)).
(For further examples see~\cref{fig:A2} in the appendix.)
For dilute samples in equilibrium, $\left \langle \rho \left(\vecc{r}, t\right) \right \rangle_t$ is related to $U(\vecc{r})$ (\cref{fig:speckles} (a)) by the Boltzmann distribution.
At mean particle densities which result in reasonable statistics, however, $\left \langle \rho \left(\vecc{r}, t\right) \right \rangle_t$ is affected by both, $U(\vecc{r})$ and particle-particle interactions.

The local time-averaged particle density $\langle\rho(\vecc{r},t)\rangle_t$ is characterized by the two-dimensional density autocovariance function $C(\vecc{r})$, i.e. the density autocorrelation function of $\langle \rho(\vecc{r},t)\rangle_t$ around the mean $\rho_0$, which is, making use of the Wiener-Khinchin theorem,~\cite{Steward2004} given by
\begin{align}
C(\vecc{r}) = & \left [\big\langle \left \langle \rho\left(\vecc{r'},t\right) \right \rangle_t \left \langle \rho\left(\vecc{r'}+\vecc{r},t\right) \right \rangle_t \big\rangle_{\vecc{r'}}\right ]_l - \rho_0^2 \nonumber\\
= &\left [\mathcal{F}^{\,\text{-}1}\left(\mathcal{F}\left\{\langle\rho(\vecc{r},t)\rangle_t\,\text{-}\,\rho_0\right\} \;\mathcal{F^*}\left\{(\langle\rho(\vecc{r},t)\rangle_t\,\text{-}\,\rho_0)\right\}\right)\right ]_l
\label{eq:CxyF1}
\end{align}
where $\mathcal{F}$, $\mathcal{F}^{-1}$, and $^*$ indicate the Fourier transformation, inverse Fourier transformation, and complex conjugation, respectively.
Since isotropic samples are considered, an azimuthal average is carried out; $C(r) = (1/2\pi)\int_{0}^{2\pi}C(r,\Theta)\text{d}\Theta$.

\begin{figure} %%%%% 555555 %%%%%
\includegraphics[width=1.0\linewidth]{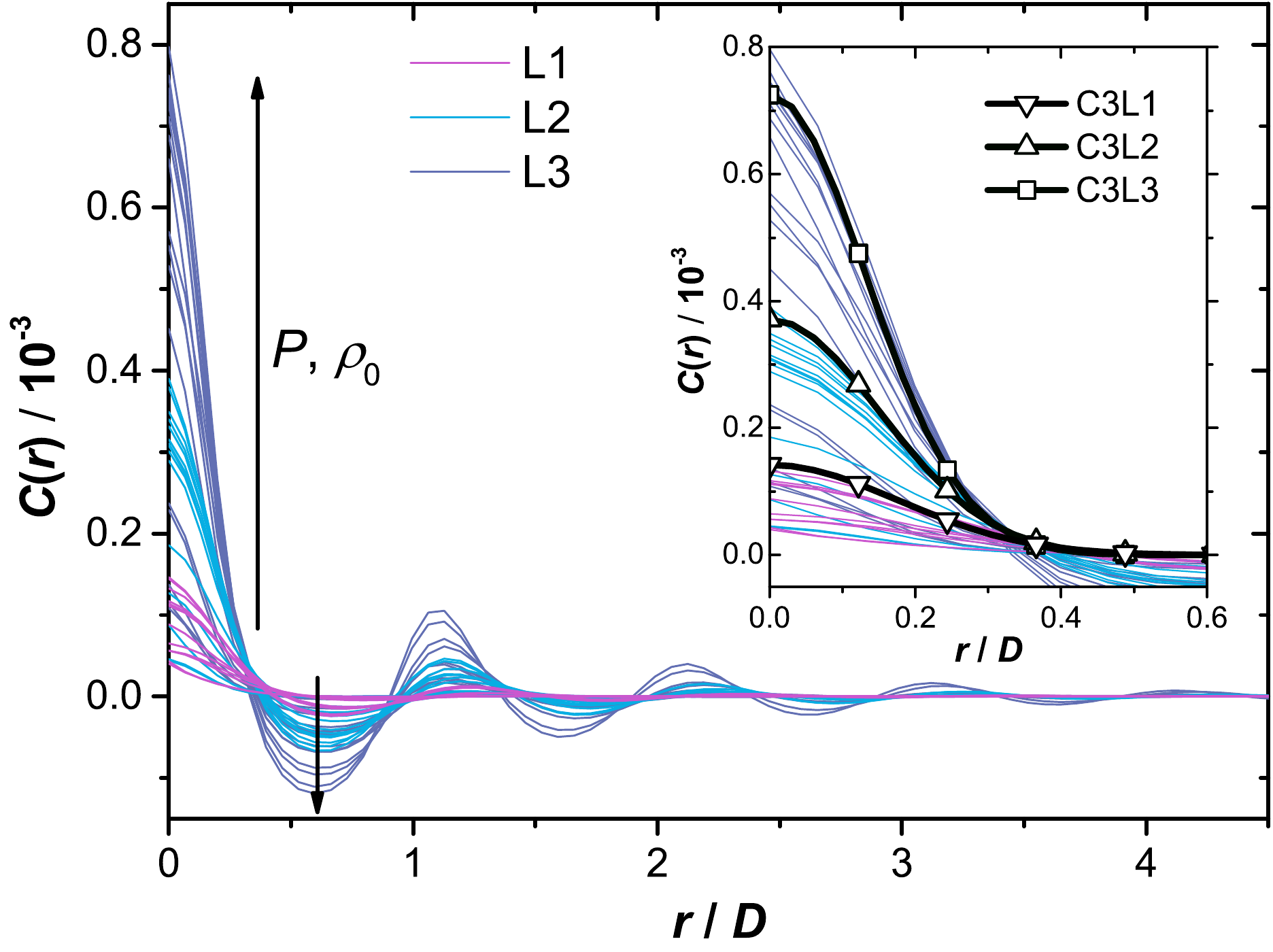}
\caption[Ccov and Corrlength]
{Azimuthally averaged autocovariance or spatial autocorrelation function $C(r)$ of the time-averaged particle density $\langle \rho(\vecc{r},t) \rangle_t$ as a function of normalized distance $r / D$ for different laser powers $P$ (L1-L3, indicated by colours) and increasing mean particle density $\rho_0$ (indicated by arrows).
Inset: Same data with Gaussian fits to data corresponding to mean particle density C3 as black lines with symbols representing different laser powers (as indicated).
\label{fig:Ccov}}
\end{figure}

\begin{figure} %%%%% 666666 %%%%%
\includegraphics[width=0.9\linewidth]{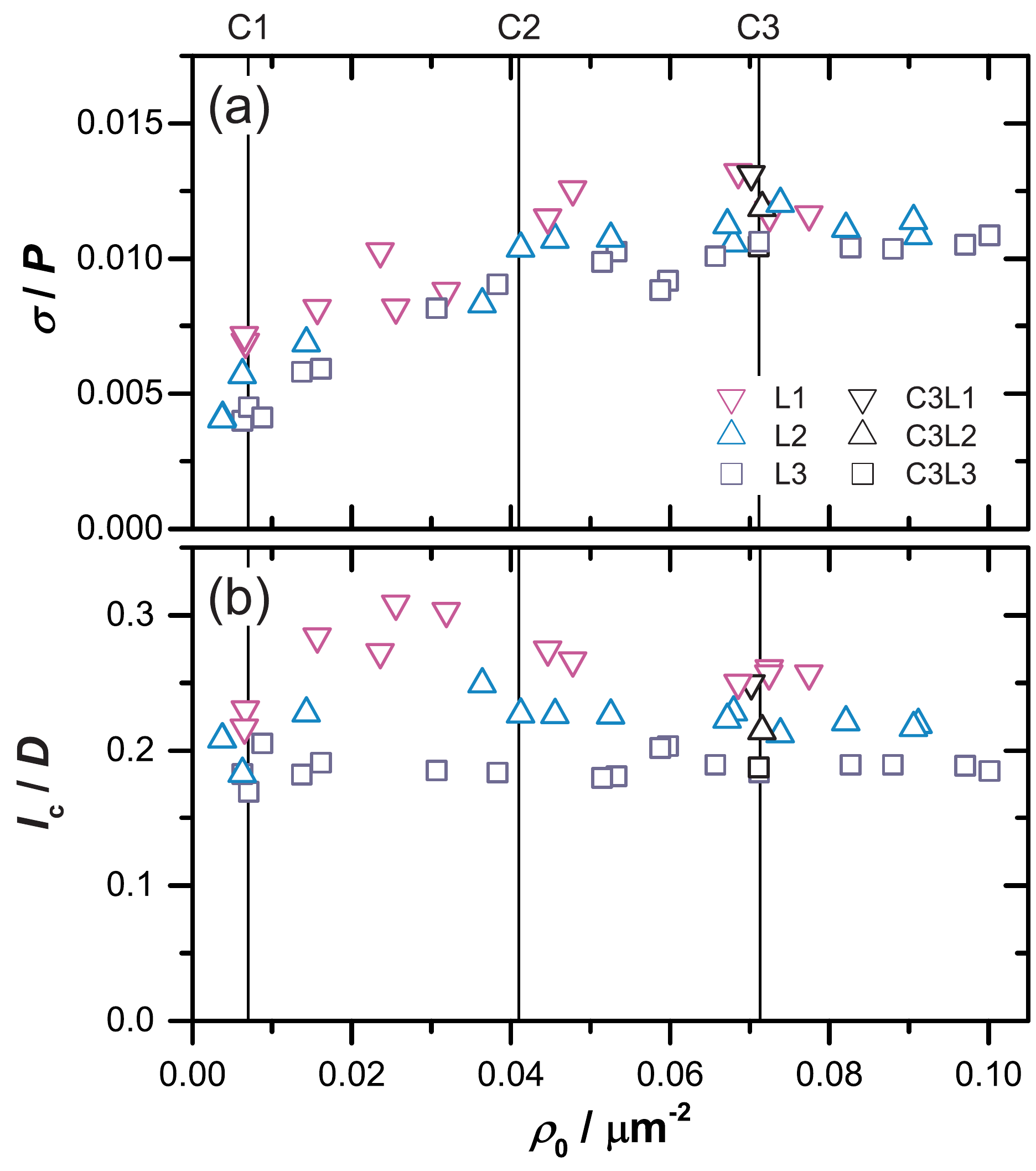}
\caption[SD]
{(a) Standard deviation $\sigma$ and (b) correlation length $l_\text{c}$ of the time-averaged particle density $\langle \rho(\vecc{r},t) \rangle_t$ as a function of mean particle density $\rho_0$ shown for different laser powers $P$ (L1-L3, as indicated).
\label{fig:SD}}
\end{figure}

In~\cref{fig:Ccov}, the azimuthally-averaged density autocovariance function $C(r)$ is shown for different laser powers.
It shows similar behaviour for all investigated experimental conditions since varying the laser power only changes the disorder strength but not the shape or statistics of the rPEL.
A pronounced peak is located at the origin which is well described by a Gaussian distribution $C(r)=\sigma^2\exp(-(r/l_\text{c})^2)$ (\cref{fig:Ccov} inset).
Its amplitude $\sigma^2 = C(0) = \langle \langle \rho (\vecc{r},t) \rangle_t^2 \rangle_{\vecc{r}} - \rho_0^2$ is the variance of the local particle density and describes the probability to find a, not necessarily the same, particle in a specific region for the entire measurement time.
Thus the amplitude $\sigma^2$ characterizes the mean depth of the potential minima as sampled by the particles.
It increases with potential strength about linearly and also increases with $\rho_0$ (\cref{fig:SD}(a)).
With increasing $\rho_0$, the particles occupy increasingly higher potential values thus broadening the range of occupied values and increasing $\sigma$.
The correlation length $l_\text{c}$ (\cref{fig:SD}(b)) characterizes the width of the potential minimum as sampled by the particles.
It decreases with laser power $P$, i.e. disorder strength, reflecting the tighter pinning.
It also depends on the mean particle density $\rho_0$.
For low $\rho_0$, particle-potential interactions dominate, whereas with increasing $\rho_0$, the particles occupy increasingly higher potential values within the same minimum and hence $l_\text{c}$ increases.
In contrast, for high $\rho_0$, particle-particle interactions dominate and the area fraction occupied by particles becomes important.
Then $l_\text{c}$ is mostly determined by the particle diameter $D$ rather than the speckle size and hence slightly decreases before reaching a constant level.
The height of this level decreases with potential strength, since the smaller the particles' excursions the smaller $l_c$.

The primary peak of $C(r)$ is followed by a minimum, which is more pronounced as the laser power increases (\cref{fig:Ccov}, indicated by arrow).
It occurs at a distance comparable to the correlation length of the potential, $0.69D$ (\cref{fig:speckles}), independent of both, $P$ and $\rho_0$.
In contrast, the minimum becomes more pronounced with $P$ and $\rho_0$.
It is caused by particles pinned in potential minima, which exclude particles from their vicinity (\cref{fig:images} (f) and (j)).
The higher order minima (and maxima) are roughly spaced by multiples of the particle diameter $D$.
These oscillations are caused by either particle-potential or, in the case of high $\rho_0$, multiple-particle interactions and thus reflect spatial arrangements of neighbouring particles, such as, e.g., caused by depletion and caging.

\subsection{Correlation Functions}
\label{sec:PPint}

To characterize the particle-potential and particle-particle interactions, based on the measured time-averaged local particle density $\langle \rho(\vecc{r},t)\rangle_t$ we determine the pair distribution or pair density correlation function $g^{(1)}(r)$, the off-diagonal density correlation function $g^{(2)}(r)$ and the total correlation or Ursell function $h(r)$ which all are normalized by $\rho_{0}^2$.

The off-diagonal density correlation function $g^{(2)}(r)$ is an analogue of the Edwards-Anderson order parameter.~\cite{Sengupta2005,Stein2013}
It is defined by
\begin{equation}
g^{(2)}(\vecc{r}) = \frac{1}{\rho_0^2} \left [\left\langle \left\langle \rho\left(\vecc{r'},t,l\right) \right\rangle_t \left\langle \rho\left({\vecc{r'}{+}\vecc{r}},t,l\right) \right\rangle_t \right\rangle_{\vecc{r'}} \right ]_l
\label{eq:g2}
\end{equation}
and hence is the normalized spatial correlation function of the mean local density among disorder realisations.
It quantifies the probability for a particle to be pinned by the rPEL, i.e. it quantifies whether a certain location is still occupied by a particle after an arbitrarily long time period.~\cite{Stein2013}
Therefore it describes a coupling between spatial disorder of pinning sites and particle positional ordering in time as well as multiple-particle interactions.
Without an external potential, i.e. vanishing disorder strength, and for low enough mean particle densities, where particle-particle interactions are not important, $g^{(2)}(\vecc{r}) = 1$.
Application of an external quenched disorder, here in the form of the speckle pattern of the external laser field, disrupts this conservation law locally and thereby breaks the corresponding symmetry.
This phenomenon is directly observed in the form of the real space inhomogeneities introduced in the density profile.
The off-diagonal density correlation function $g^{(2)}(\vecc{r})$ characterizes the order parameter of this symmetry-broken disordered state.
Furthermore, for a large field of view and hence disorder averaging in one single realisation of the rPEL, $g^{(2)}(\vecc{r}) = C(\vecc{r}) / \rho_0^2 +1$.
We consider the azimuthal average $g^{(2)}(r)$.
It is calculated from the experimental data by
\begin{align}
g^{(2)}(r) = &\frac{1}{\rho_0^2} \frac{1}{L} \sum_{l=1}^L \; \frac{1}{MN} \sum_{m',n'=1}^{MN}\nonumber\\
&\times\frac{1}{N_\text{r}}\sum_{m,n} \; \left [ \left \{ \frac{1}{K} \sum_{k=1}^K \rho(x_{m'},y_{n'},t,l) \right \}\right.\nonumber\\
&\left. \times \left \{ \frac{1}{K} \sum_{k=1}^K\rho(x_{m'+m},y_{n'+n},t,l) \right \} \right ]
\label{eq:g2_2}
\end{align}
where $m$ and $n$ are chosen such that regions with their centres in an annulus between radii $r{-}\Delta r/2$ and $r{+}\Delta r/2$ are included with $N_\text{r}$ the number of such regions.

\begin{figure} %%%%% 777777 %%%%%
\includegraphics[width=0.9\linewidth]{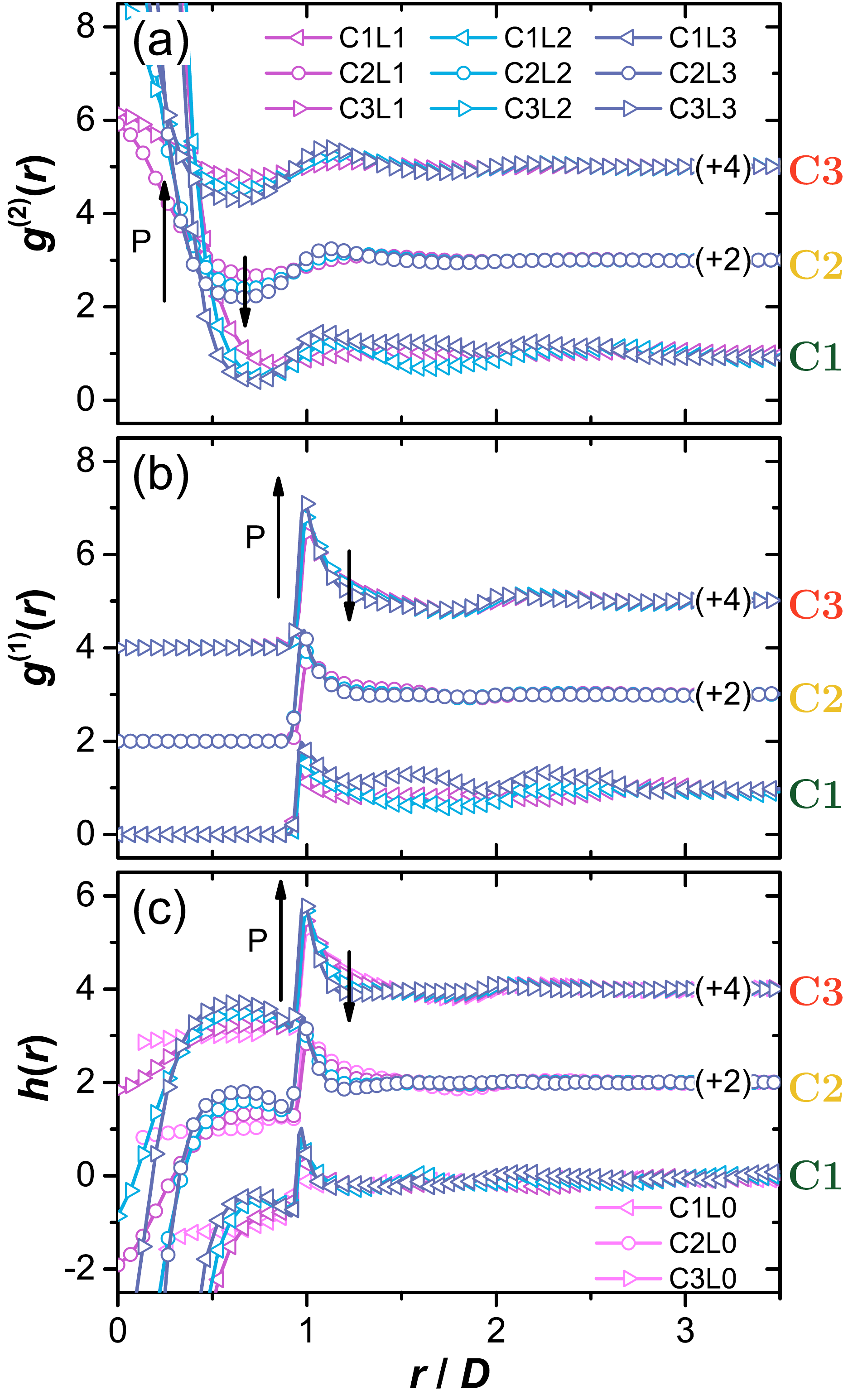}
\caption[g1rg2rhr_3x3]
{(a) Azimuthally averaged off-diagonal density correlation function $g^{(2)}(r)$, (b) pair density correlation function $g^{(1)}(r)$, and (c) total correlation function $h(r)$ as a function of the normalized distance $r / D$ for different laser powers $P$ (L1-L3, indicated by arrows) and mean particle densities $\rho_0$ (C1-C3, as indicated).
The data corresponding to C2 and C3 were shifted along the y-axis by $+2$ and $+4$, respectively.
\label{fig:g1rg2rhr_3x3}}
\end{figure}

\Cref{fig:g1rg2rhr_3x3} (a) shows $g^{(2)}(r)$ for different mean particle densities $\rho_0$ (C1-C3) and laser powers $P$ (L1-L3, indicated by arrows).
(Further conditions are shown in~\cref{fig:A3} in the appendix.)
For large distances $r$ the time-averaged particle density is uncorrelated and thus $g^{(2)}(r\to\infty)=1$.
By contrast, correlations between high local densities, reflecting potential minima, lead to deviations from unity.
For small distances $r\to0$ a pronounced peak is observed, consistent with the observations in connection with the density autocovariance function $C(r)$ (cf.~\cref{fig:Ccov}).
For distances larger than the minimal particle-particle distance $r>D$, no clear $r$ dependence of the fluctuations is visible for the lowest $\rho_0$ (C1).
This is attributed to the irregular distribution of the small number of particles in the random potential, in particular the potential minima, and hence the limited sampling (see~\cref{sec:sampling}).
For medium and high $\rho_0$ (C2, C3) maxima occur around multiple integers of $D$.
In the absence of a rPEL no such fluctuations are present in $g^{(2)}(r)$ (\cref{fig:A3} in the appendix).
This indicates the interplay of particle-particle and particle-potential interactions. 

The correlation function $g^{(1)}(\vecc{r})$, which is the disorder-averaged analogue of the pair distribution function or pair density correlation function, is defined by~\cite{Sengupta2005}
\begin{equation}
g^{(1)}(\vecc{r},l) = \frac{1}{\rho_0^2} \left[\left\langle \rho\left(\vecc{r'},t,l\right) \rho\left({\vecc{r'}{+}\vecc{r}},t,l\right) \right\rangle_{t,\vecc{r'}} \right]_l  - \frac{1}{\rho_0}\delta(\vecc{r},l)
\label{eq:g1}
\end{equation}
where $\delta(\vecc{r},l)$ is the Dirac delta function and the time average for the disordered system has to be taken prior to the disorder average.
Note that the time-average of the product of the densities is taken in~\cref{eq:g1}, whereas the product of the time-averaged densities is considered in~\cref{eq:g2}.
In the canonical ensemble the last term vanishes.
The azimuthal average can be determined from the experimental data by
\begin{align}
g^{(1)}(r) = &\frac{1}{\rho_0^2} \frac{1}{L} \sum_{l=1}^L \; \frac{1}{MN} \sum_{m',n'=1}^{MN} \; \frac{1}{N_\text{r}} \sum_{m,n}\nonumber\\
&\times \frac{1}{K} \sum_{k=1}^K \rho(x_{m'},y_{n'},t,l) \rho(x_{m'+m},y_{n'+n},t,l)
\label{eq:g1_sum}
\end{align}
where, again, $m$ and $n$ are chosen to include regions with their centres in an annulus between radii $r{-}\Delta r/2$ and $r{+}\Delta r/2$.
It describes the spatial variance in the time-averaged local particle density.~\cite{Hansen2006}

For $r<D$, $g^{(1)}(r)=0$ whereas $g^{(1)}(r)=1$ for $r\gg D$ for all conditions (\cref{fig:g1rg2rhr_3x3}~(b)), which resembles a hard sphere system.
At intermediate $r$, oscillations similar to the ones found for $g^{(2)}(r)$ are observed.
For large $\rho_0$ they hardly depend on the laser power $P$.
At low $\rho_0$ the fluctuations are more pronounced but appear at random distances.
This is attributed to the limited sampling of the rPEL due to the small number of particles (see~\cref{sec:sampling}).

\begin{figure} %%%%% 888888 %%%%%
\includegraphics[width=0.9\linewidth]{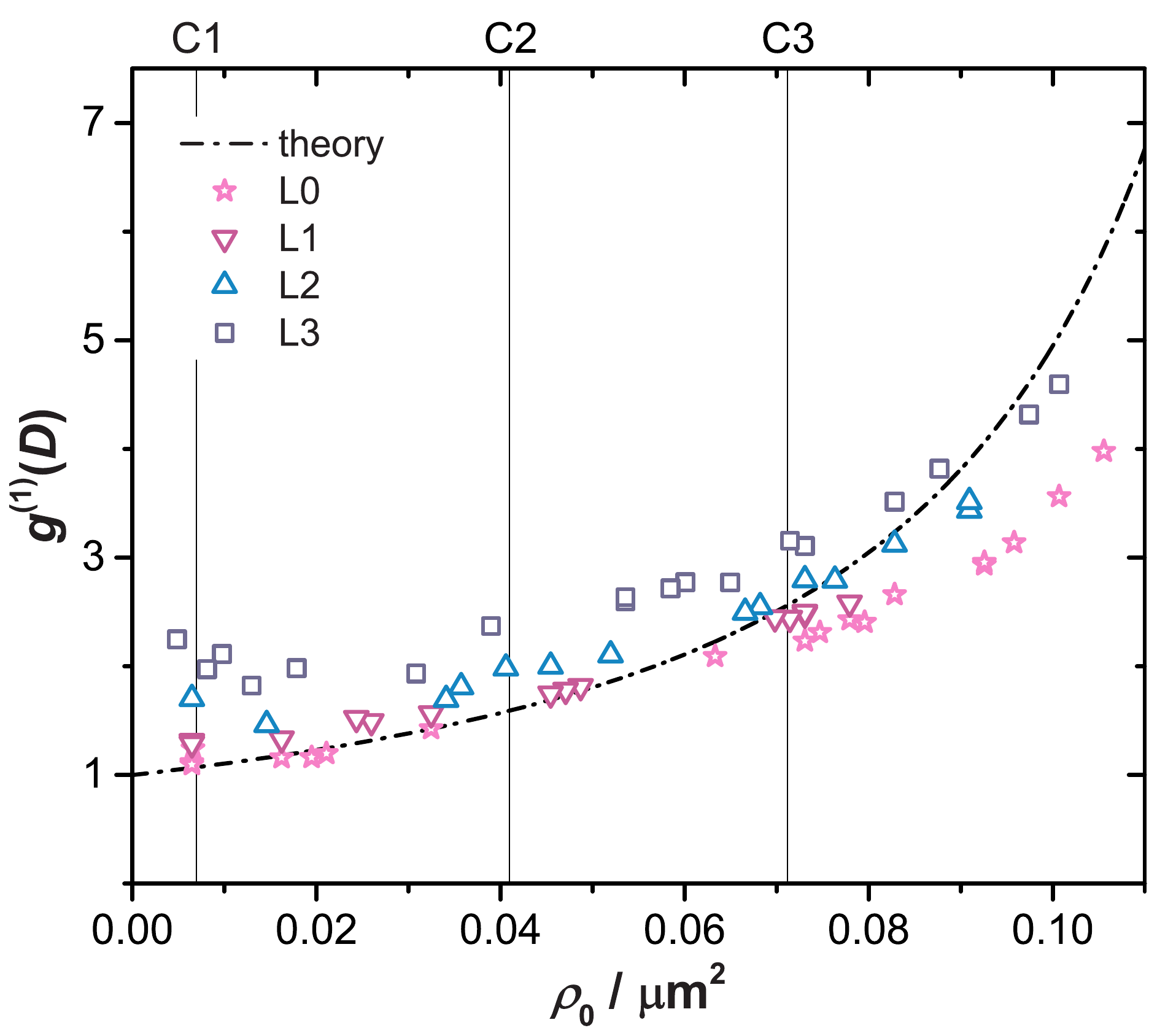}
\caption[$g^(1)(D)$]
{Contact value of the pair density correlation function $g^{(1)}(D)$ as a function of the mean particle density $\rho_0$ for different laser powers $P$ (L0-L3, as indicated).
The dashed line represents the prediction by the Henderson equation of state.~\cite{Henderson1975}
\label{fig:g1peak}}
\end{figure}

The peak at $r=D$, the contact value $g^{(1)}(D)$, is linked to the compressibility and thus the equation of state~\cite{Helfand1961,Henderson1975,Guo2006} (\cref{fig:g1peak}).
The contact value $g^{(1)}(D)$ increases with $\rho_0$ and $P$.
The experimentally determined $g^{(1)}(D)$ is very sensitive to the number of particles and their localization errors as well as the histogram parameters, i.e. bin positions and size.
In particular at higher densities ($\rho_0 > 0.06~\upmu \text{m}^{-2}$), the peak of $ g^{(1)}(r)$ at $r\approx D$ is very sharp compared to the bin size and the uncertainty of our tracking procedure and therefore $g^{(1)}(D)$ is expected to be underestimated.
A theoretical prediction for hard spheres,~\cite{Henderson1975,Guo2006} $g^{(1)}(D) =\left(1-7\phi_\text{A}/16\right)/\left(1-\phi_\text{A}\right)^2$ (\cref{fig:g1peak}) agrees with the experimental data obtained in the absence of a rPEL (L0, indicated by pink stars) for low densities ($\rho_0\lesssim 0.06~\upmu \text{m}^{-2}$) but differs at higher densities.
This is possibly caused by the above mentioned uncertainties involved in the determination of $g^{(1)}(D)$.

The total correlation or Ursell function $h(r)$ is given by
\begin{align}
h(r) &= g^{(1)}(r) - g^{(2)}(r) \; \text{.}
\label{eq:h}
\end{align}
The contributions of particle-potential interactions to $g^{(1)}(r)$ are taken into account by $g^{(2)}(r)$ and hence $h(r)$ mainly describes the disorder-, ensemble- and time-averaged density fluctuations caused by particle-particle and multiple-particle interactions.
Therefore, $h(r)$ appears as a pair distribution function which hardly contains correlations due to the potential, in particular for $r>D$.
For a homogeneous, isotropic fluid in the absence of an external potential, and hence $g^{(2)}(r)=1$, it becomes $h(r) = g^{(1)}(r) - 1$, resembling the pair correlation function.

\begin{figure} %%%%% 999999 %%%%%
\includegraphics[width=0.85\linewidth]{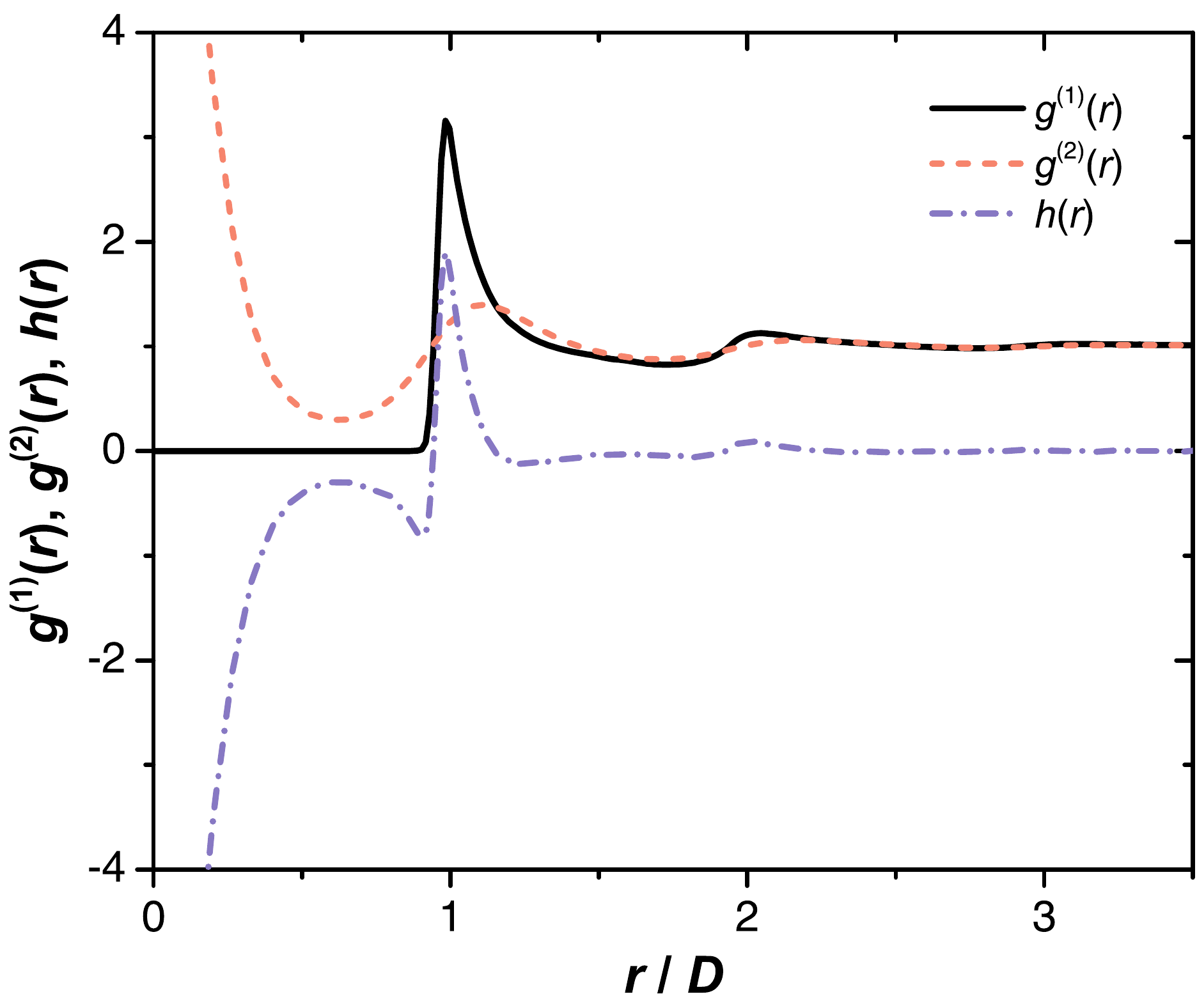}
\caption[g1g2h]
{Comparison of the azimuthally averaged pair density correlation function $g^{(1)}(r)$, off-diagonal density correlation function $g^{(2)}(r)$, and total correlation or Ursell function $h(r)$ as a function of normalized distance $r / D$ for high laser power L3 and mean particle density C3.
\label{fig:g1g2h}}
\end{figure}

The total correlation function $h(r)$ is shown in \cref{fig:g1rg2rhr_3x3} (c) for different mean potential densities $\rho_0$ (C1-C3) and laser powers $P$ (L0-L3).
In the absence of a rPEL (L0), $h(r)$ is approximately -1 for $r<D$, shows a peak at $r\approx D$ and is about zero beyond the peak for $r\gg D$.
In the presence of a rPEL, the behaviour for $r<D$ differs due to the strongly increasing $g^{(2)}(r)$.
The height of the peak at $r\approx D$ increases with increasing mean particle density and its width decreases with increasing laser power.
Remarkably, beyond this peak $h(r)$ is almost constant and takes a value of about zero for all investigated mean particle densities and laser powers.
This is due to the balance between $g^{(1)}(r)$ and $g^{(2)}(r)$ which is illustrated in~\cref{fig:g1g2h} by a direct comparison of all three functions.
The above-mentioned concurrence of the oscillations of $g^{(1)}(r)$ and $g^{(2)}(r)$ results in an almost flat $h(r)$ beyond the first peak.
The remaining maximum of $h(r)$ at $r\approx 2D$ is rather attributed to particle-particle and multiple-particle interactions than particle-potential interactions.
(For a comparison of $g^{(1)}(r)$ and $g^{(2)}(r)$ at all measured combinations of mean particle density $\rho_0$ and laser power $P$ see~\cref{fig:A3} in the appendix.)

\subsection{Replica Liquid State Theory}
\label{sec:replica}

For a deeper understanding of our results, we compare the experimentally obtained correlation functions $g^{(1)}(r)$ and $g^{(2)}(r)$ to predictions of liquid state theory,~\cite{Hansen2006} generalised to include the effects of an external rPEL, i.e. quenched disorder.
While the details of this theory have been described previously,~\cite{Menon1994, Sengupta2005} they are briefly mentioned for completeness. 

The colloidal particles are assumed to interact with each other through a hard sphere pair potential $V(r)$ and are exposed to a random potential $U({\bf r})$ with the distribution of energy values $p(U)$ being Gaussian and the short ranged spatial correlations quantified by $C_{U}(r)$ as in the experiments.
To obtain the free energy of this system, the disorder-average of the logarithm of the partition function, $[\ln Z]_l$, is calculated using the replica trick,~\cite{Edwards1975}
$$[\ln Z]_l = \lim_{q \to 0} \int dU p(U) \frac{Z^q - 1}{q} \; \text{,}$$ where $Z^q$ is the partition function of a set of $q$ non-interacting realisations of the same system, i.e.~`replicas'.
The partition function of $\mathcal{N}$ particles replicated $q$ times and averaged over the disorder distribution $p(U)$ is identical to the partition function of $\mathcal{N} \times q$ particles interacting with the potential $V^{\alpha \beta}(r) = V(r)\delta_{\alpha \beta} + C_U(r)$.~\cite{Menon1994}
The liquid state theory for such a system is now constructed assuming replica symmetry where all liquid state correlation functions, such as the pair correlation function, share the symmetry $g^{\alpha \beta}(r) = g^{\beta \alpha}(r) = g^{(1)}(r)\delta_{\alpha \beta} + g^{(2)}(r)(1 - \delta_{\alpha \beta})$.
In the $q \to 0$ limit, the Ornstein-Zernike relation is~\cite{Hansen2006}
\begin{eqnarray}
h^{(1)}(k) & = & \frac{c^{(1)}(k) - (c^{(1)}(k) - c^{(2)}(k))^2}{(1 - c^{(1)}(k) + c^{(2)}(k))^2} \nonumber \\
h^{(2)}(k) & = & \frac{c^{(2)}(k)}{(1 - c^{(1)}(k) + c^{(2)}(k))^2} \; \text{,}
\end{eqnarray}
where $h^{(1)}(k)$ is the Fourier transform of the (diagonal) pair correlation function $h^{(1)}(r) = g^{(1)}(r) - 1$ and $c^{(1)}(r)$ the corresponding direct correlation function.
The off-diagonal correlations, with superscript $(2)$, are defined analogously.
The Ornstein-Zernike relation needs to be complemented with a closure relation in order to solve for the correlation functions.
We have used two sets of closure relations to try to reproduce the measured correlation functions.
Firstly, the analogue of the Percus-Yevick (PY) equation modified for the replicated case, 
\begin{align}
c^{(1)}(r) &= \left(e^{-\beta (V(r) + C_U(r))} - 1\right)\left(1 + y^{(1)}(r)\right) \nonumber \\
c^{(2)}(r) &= \left(e^{-\beta C_U(r)} - 1\right)\left(1 + y^{(2)}(r)\right) \; \text{,}
\label{eq:PY}
\end{align}  
where $y^{(1)}(r) = h^{(1)}(r) - c^{(1)}(r)$ and similarly $y^{(2)}(r)$ are the indirect correlation functions.
These relations are solved using the method of Gillan.~\cite{Gillan1979} 

\begin{figure} %%%%% 101010 %%%%%
\includegraphics[width=1.00\linewidth]{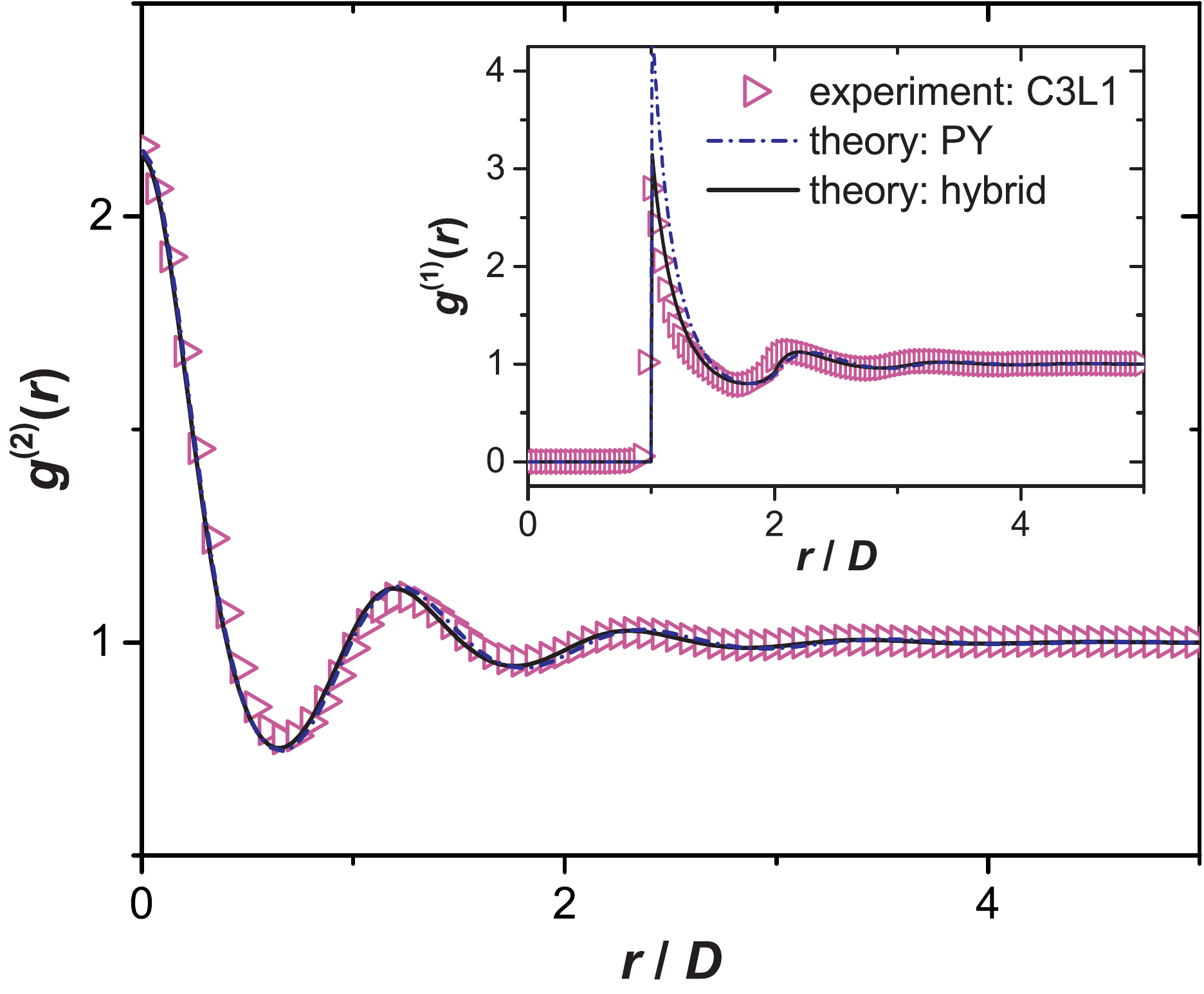}
\caption[g1g2h]
{Comparison of the experimentally determined azimuthally averaged off-diagonal density correlation function $g^{(2)}(r)$ and the pair density correlation function $g^{(1)}(r)$ (inset) with results obtained from liquid state theory, as a function of normalized distance $r / D$ for low laser power L1 and high mean particle density C3.
\label{fig:g1g2theory}}
\end{figure}

The results from the replicated PY liquid state theory are compared to the experimental results for C3L1, i.e. a mean particle density $\rho_0=0.56\,D^{-2}$ (\cref{fig:g1g2theory}).
Fitting yielded for the strength of the disorder $\langle U^2\rangle^{1/2} = 1.8 \, k_\text{B} T$ with the thermal energy $k_\text{B} T$, consistent with experimental expectations, and for the correlation length $\xi = 0.43\,D$, which is somewhat lower than the experimental value $\xi = 0.69\,D$.
While the $g^{(2)}(r)$ agree remarkably well, the PY approximation overestimates correlations in $g^{(1)}(r)$.
This is a well known feature of the PY closure.
To correct for this, we propose and solve a hybrid set of closure relations where the first equation of the set in~\cref{eq:PY} is replaced with 
\begin{eqnarray}
c^{(1)}(r) & = & e^{-\beta (V(r) + C_U(r)) + y^{(1)}(r)} - 1 - y^{(1)}(r)
\label{eq:hybrid}
\end{eqnarray}  
and the second equation is kept the same.
This results in much better agreement of the $g^{(1)}(r)$ while the $g^{(2)}(r)$ is almost unchanged. 
Thus, with the hybrid set of closure relations quantitative agreement between experimental data and replica liquid state theory predictions are obtained.

For experiments with the same laser power $P$, also the strength of the disorder $\langle U^2\rangle^{1/2}$ and the correlation length $\xi$ remain constant, independent of the mean particle density $\rho_0$.
Ideally, the results from our replica liquid state theory should follow these expectations. However, at large laser powers Eqs. (10) and (11) begin to give unphysical results. Also the fitted values, especially for  $\xi$, depend on $\rho_0$. This indicates that the validity of the simple closure relations used in our theory is limited if the disorder is strong. Moreover, it is important to ensure that the whole landscape is sampled by the particles, which is particularly difficult for dilute systems within a reasonable measurement time.
This can only be resolved by further experiments on a larger set of densities $\rho_0$ and laser powers $P$ and/or by a better liquid state theory.~\cite{Rogers1984}

Finally, the time-averaged local particle density in the presence of the rPEL is given by:~\cite{Hansen2006,Sengupta2005}
\begin{equation}
\langle \rho(\vecc{r},t,l) \rangle_t = \rho_0 - \frac{\rho_0^2}{k_\text{B}T} \int d{\bf r}' h(|{\bf r} - {\bf r}'|)U({\bf r}') + \text{...} %- \frac{\rho_0}{k_\text{B}T} U({\bf r}) + \text{...}
\end{equation}
which links the time-averaged local particle density $\langle \rho(\vecc{r},t,l) \rangle_t$ to the disorder potential $U(\vecc{r})$.
This analytical relationship can be used to determine $U(\vecc{r})$ from a measurement of $\langle \rho(\vecc{r},t,l) \rangle_t$ or to predict $\langle \rho(\vecc{r},t,l) \rangle_t$ from $U(\vecc{r})$ and $h(r)$.~\cite{Sengupta2005}

%%%%% Conclusions %%%%%
\section{Conclusions}
\label{sec:conclusions}

We investigated colloidal particles in a random potential energy landscape (rPEL) with energy values distributed according to a Gamma distribution.
It was imposed by a laser speckle pattern.
The rPEL affects the distribution of particles which, at higher mean particle densities, is also modified by particle-particle interactions.
Therefore, local particle density variations occur, which are correlated in time and space.
The time-averaged local particle density was determined and analysed as a function of mean particle density $\rho_0$ and laser power $P$, i.e. disorder strength.
The off-diagonal density correlation function $g^{(2)}(r)$ not only reflects the potential roughness, but also spatial correlations in the local density caused by pinned particles.
Thus it reflects particle-potential and particle-particle interactions.
The pair density correlation function $g^{(1)}(r)$ is also influenced by spatial correlations of the rPEL.
As a result, the total correlation or Ursell function $h(r)=g^{(1)}(r)-g^{(2)}(r)$ hardly reflects particle-potential interactions, but characterizes particle-particle and multiple-particle interactions.
To our knowledge, this is the first time these correlation functions have experimentally been determined in the presence of disorder.
Furthermore, they have successfully been compared to results from replica liquid state theory.
This results in quantitative agreement, but also points towards deficits in the existing liquid state theory and calls for further experiments.

%%%%% Acknowledgements %%%%%
\section*{Acknowledgments}

We thank Manuel Escobedo-Sanchez, J\"urgen Horbach, Hartmut L\"owen, Gautam Menon, and Christoph Zunke for very helpful discussions and suggestions.
We acknowledge funding by the Deutsche Forschungsgemeinschaft (DFG) (grant EG269/6-1) and the FP7-PEOPLE-2013-IRSES (grant no: 612707, DIONICOS).

%\clearpage
% % % Appendix % % %
% or use
%\onecolumngrid
%\appendix
\begin{figure*}%%%%% A1A1A1 %%%%%
\includegraphics[width=1.0\linewidth]{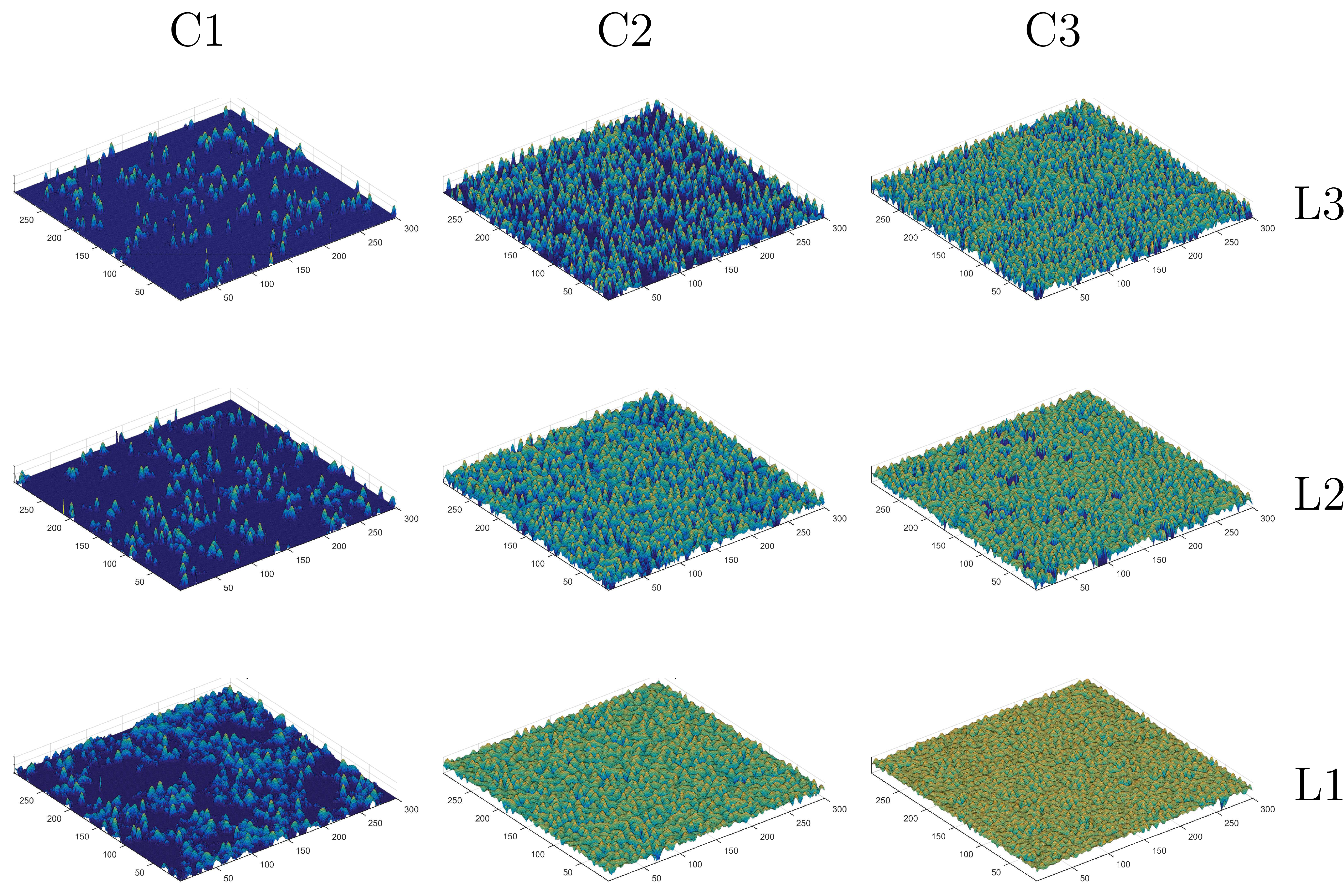}
\caption[rhoxy]
{Time-averaged local particle density $\langle \rho(\vecc{r}, t) \rangle_t$ for increasing laser power (L1-L3) and mean particle density (C1-C3, as indicated).
Colour scale indicates low to high densities by blue to red colours, where different scales are used for the different conditions.
\label{fig:A2}}
\end{figure*}

%%%%% References %%%%%
\bibliography{Potential_arxive}

\section*{Appendix}
\subsubsection{Particle Arrangements}
\begin{figure}[h]%%%%% A2A2A2 %%%%%
\includegraphics[width=1.0\linewidth]{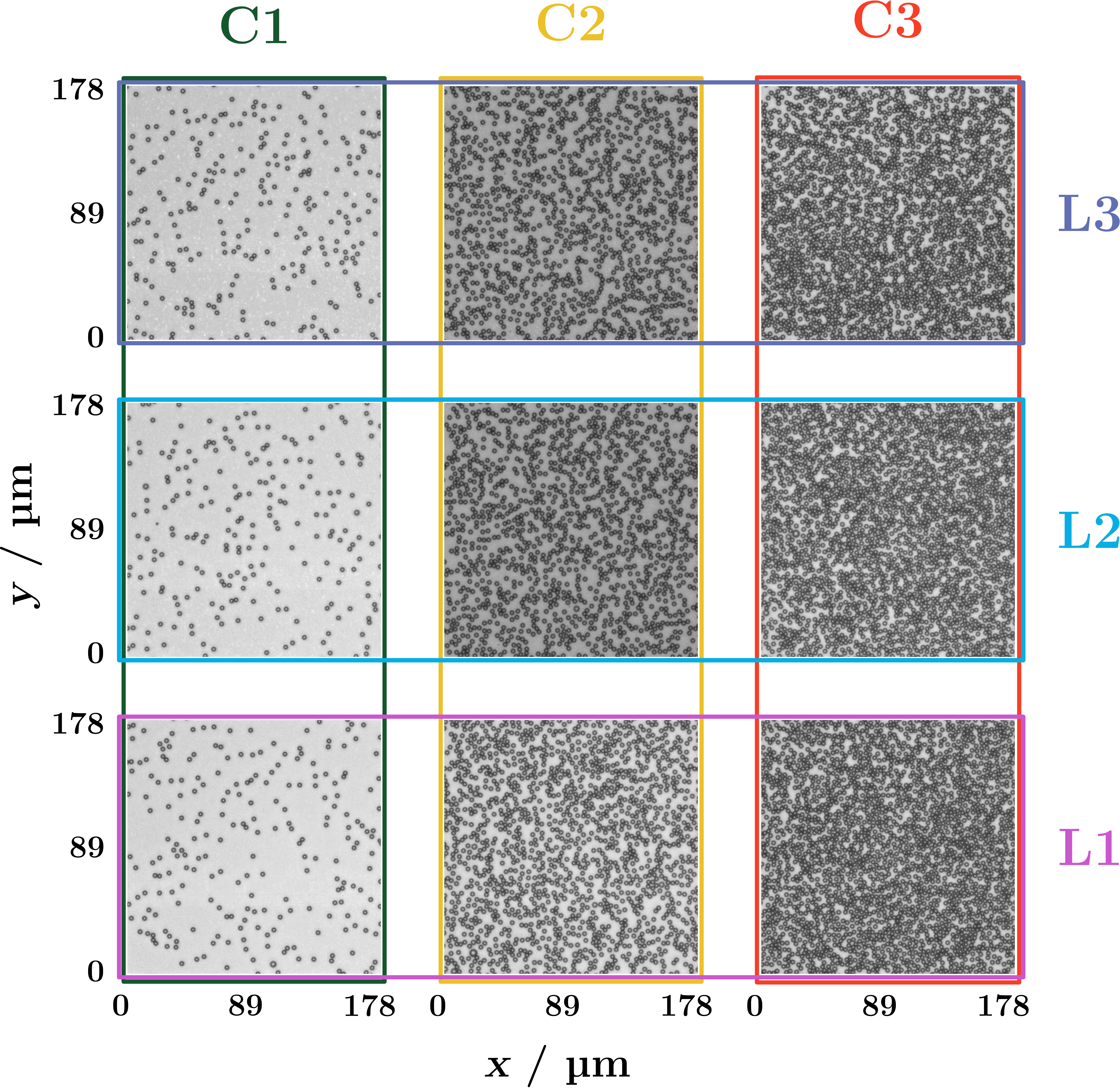}
\caption[images]
{Micrographs of parts of the samples ($178\times178~\upmu\text{m}^2$) for increasing laser power (L1-L3) and mean particle density (C1-C3, as indicated).
\label{fig:A1}}
\end{figure}

\Cref{fig:A1} shows micrographs of colloidal particles for three different mean particle densities $\rho_0$ (C1-C3) and increasing laser power $P$, i.e. disorder strength, (L1-L3).
Neither for low nor for high mean particle density and/or laser power an effect of the potential is immediately visible in the images.

\subsubsection{Time-Averaged Local Particle Density}
The time-averaged local particle density $\left \langle \rho \left(\vecc{r}, t\right) \right \rangle_t$ for three different laser powers $P$, i.e. disorder strengths, (L1-L3) and mean particle densities $\rho_0$ (C1-C3) is shown in~\cref{fig:A2}.
For dilute samples trapping of particles in deep potential minima during the entire measurement time leads to a discretisation of the density landscape.
This becomes stronger with increasing laser power.
At higher mean particle densities, $\left \langle \rho \left(\vecc{r}, t\right) \right \rangle_t$ is affected by both particle-potential and particle-particle interactions, resulting in a smoothed density landscape.
This becomes more apparent with a decrease in the laser power.

\subsubsection{Correlation Functions}
\begin{figure}%%%%% A3A3A3 %%%%%
\includegraphics[width=1\linewidth]{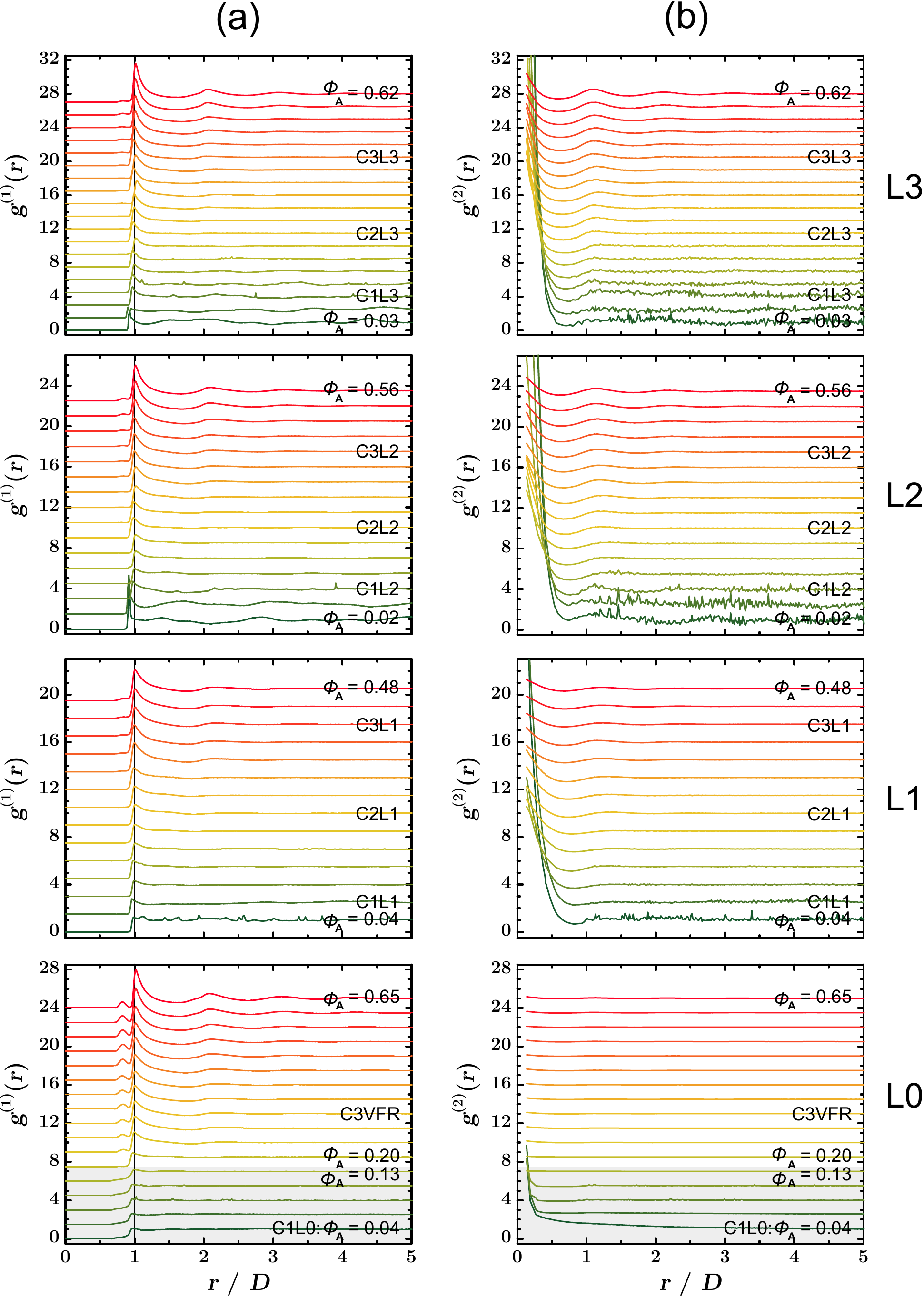}
\caption[images]
{(a) Azimuthally-averaged pair density correlation function $g^{(1)}(r)$ and (b) off-diagonal density correlation function $g^{(2)}(r)$ for increasing laser power (L0-L3) and mean surface fraction $\phi_\text{A}$ or particle density (C1-C3, as indicated by colour gradient from green to red).
Data are shifted vertically for clarity.
\label{fig:A3}}
\end{figure}

The azimuthally-averaged pair density correlation function  $g^{(1)}(r)$ and off-diagonal density correlation function $g^{(2)}(r)$ at all measured combinations of mean particle density $\rho_0$ and laser power $P$ (L0-L3) are shown in~\cref{fig:A3}.
For very large distances $r$ the time-averaged local particle density is uncorrelated, and thus $g^{(1)}(r\to\infty)=1$ and $g^{(2)}(r\to\infty)=1$ independent of the mean particle density $\rho_0$ and the laser power $P$, i.e disorder strength.
By contrast, correlations at finite distances $r$ between high local density values reflect pinning sites, i.e. particle cages or potential minima, and can be identified by deviations from this value.
In the absence of a rPEL (L0), $g^{(1)}(r)$ shows a strong dependence on the mean particle density whereas $g^{(2)}(r)\approx 1$ for all mean particle densities, except for very few low mean particle densities $\rho_0$ which is attributed to insufficient statistics.
However, in the presence of a rPEL (L1-L3) and for medium to high mean particle densities $\rho_0$, for both correlation functions maxima are observed around integer multiples of $D$, which increase with mean particle density $\rho_0$ and laser power $P$ and indicate the interplay of particle-particle and particle-potential interactions.

\end{document}